\documentclass[runningheads]{llncs}
\usepackage[nointegrals]{wasysym}
\usepackage{graphicx}
\usepackage{booktabs}
\usepackage{url}
\usepackage{tabularx}
\usepackage{floatrow}

\begin{document}
\title{Pakistan's Internet Voting Experiment}

\titlerunning{Pakistan's Internet Voting Experiment}
%
\author{Hina Binte Haq \inst{1}  \and Ronan McDermott \inst{2}   \and
Syed Taha Ali \inst{1}}

\authorrunning{HB Haq, R McDermott et al.}

\institute{School of Electrical Engineering and Computer Sciences (SEECS), National University of Sciences and Technology (NUST), Islamabad, Pakistan. \email{\{hhaq.dphd18seecs, taha.ali\}@seecs.edu.pk} \and MCDIS \email{ronan@mcdis.com}}

\maketitle 

\begin{abstract}

Pakistan recently conducted small-scale trials of a remote Internet voting system for overseas citizens. In this contribution, we report on the experience: we document the unique combination of socio-political, legal, and institutional factors motivating this exercise. We describe the system and it's reported vulnerabilities, and we also highlight new issues pertaining to materiality. If this system is deployed in the next general elections \textemdash as seems likely \textemdash this development would constitute the largest enfranchised diaspora in the world. Our goal in this paper, therefore, is to provide comprehensive insight into Pakistan's experiment with Internet voting, emphasize outstanding challenges, and identify directions for future research.

\end{abstract}

\centerline {This paper has been accepted for publication in TUT Press proceedings of }
\centerline{E-Vote-ID 2019. E-Voting.CC GmbH is the owner of the copyright.}

\section{Introduction}
\label{sec:introduction}

Pakistan recently piloted a remote Internet voting system for overseas citizens. This system, \emph{i-Voting}\footnote{also referred to as iVoting, iVOTE, IVoting}, was indigenously developed and originally scheduled for full-scale deployment in the General Elections of July, 2018. However, these plans were deferred after a third-party technical audit of the system uncovered numerous vulnerabilities and security concerns. i-Voting was deployed shortly after on a trial basis: in bye-elections, first in October, 2018, spanning 35 constituencies, and next in December, 2018 in 1 constituency. Some 7,538 votes were cast (7,461 in October and 77 in December) using this system and these were declared binding and incorporated into the final results.

It is widely expected that this pilot is a prelude to full-scale deployment in the General Elections of 2023.\footnote{Currently, the law restricts use of such systems to bye-elections.} Since Pakistan currently has over 8 million overseas citizens  \cite{overseaspopulation}, this may well be the largest deployment of Internet voting in the world. It is therefore essential to document and study this experiment.

In this paper, we make the following contributions:
\begin{enumerate}
    \item We report on the deployment: we document the public debate on Internet voting and the legislative and political process to facilitate it. We describe the i-Voting system and we report on the pilot exercise.
    
    \item We describe the various risks involved in this modality of voting and summarize key findings of the technical audit. We examine the materiality, and therefore the potential political significance of overseas voting.
    
    \item We highlight the unorthodox combination of unique political and social factors in Pakistan that have resulted in this exercise and we discuss various particular challenges that may arise as a result.
\end{enumerate}

This paper holds relevance for election stakeholders including governments, political parties, election administrators, political scientists, researchers, and technologists. Pakistan's Internet voting experience may also prove instructive for other countries, particularly in the developing world, where governments are severely limited in terms of financial resources, technical expertise, and infrastructure to undertake such critical large-scale projects. \section{Background}
\label{sec:background}

\textbf{Organization of Government}
Pakistan has a parliamentary form of government with bicameral legislature, comprising a Senate (upper house) with 104 members and a National Assembly (lower house) with 342 members. Each of the four large provinces have a unicameral legislature, consisting of a Provincial Assembly.\footnote{Federally Administered Tribal Areas (FATA) and Federal Capital Islamabad are administrative divisions in addition to the four provinces, included in the contested elections and comprise National Assembly seats only.} The electoral system is the first-past-the-post system under universal adult suffrage. Members of the National Assembly and Provincial Assembly are elected by representation in electoral districts (referred to as seats or \emph{constituencies}). The number of seats in each administrative division is listed in Table ~\ref{tab:orgl}.  General elections are conducted every five years and are overseen by the Election Commission of Pakistan (ECP), which is an independent and autonomous body as defined in the Constitution of Pakistan.  

\vspace{-4mm}
\begin{table*}[hb]
{\tiny
  \centering
    \begin{tabular*}{1\textwidth}{@{\extracolsep{\fill}}cccccccc}
    \toprule
    \textbf{Body} & \textbf{Total} & \textbf{Federal} & \textbf{Baluch-} & \textbf{Federally} & \textbf{Khyber} & \textbf{Punjab} & \textbf{Sindh } \\
   \textbf{} & \textbf{Seats} & \textbf{Capital} & \textbf{-istan} & \textbf{Administered} & \textbf{Pakhtun-} & \textbf{} & \textbf{} \\
    \textbf{ } & \textbf{ } & \textbf{Islamabad} & \textbf{} & \textbf{Tribal Areas} & \textbf{-khwa} & \textbf{ } & \textbf{ } \\
    
    \midrule
   \textbf{National} & 272   & 3     & 16    & 12    & 39    & 141   & 61 \\
   \textbf{Assembly} &    &      &     &     &     &    &  \\
    \midrule
   \textbf{Provincial} & 577   & -     & 51    & -     & 99    & 297   & 130 \\
    \textbf{Assembly} &    &      &     &     &     &    &  \\
    \bottomrule
    \end{tabular*}%
    \label{tab:orgl}%
    \vspace{-2mm}
  \caption{National and Provincial Assembly Seats \cite{constituencies}}}
\end{table*}%
\vspace{-6mm}

\noindent \textbf{Overseas Pakistanis and the Right to Vote}
Pakistan, has over 8 million overseas citizens \cite{overseaspopulation} which comprises the sixth largest diaspora in the world \cite{migrationdata}. Overseas citizens are actively engaged in the socioeconomic well-being of the country and every year send home remittances worth approximately US\$19 billion, which accounts for around 5\% of Pakistan's GDP \cite{remittances}.

Article 17 of the Constitution of Pakistan grants all adult citizens the fundamental right to vote \cite{article17}. This article has generally been interpreted to acknowledge that this right extends to all Pakistani citizens, irrespective of place of residence. Overseas Pakistanis have raised calls for enfranchisement and facilitation of their voting rights since the first general elections of 1970 \cite{dawnpildat}.

The earliest constitutional petition filed to facilitate overseas voters was in 1993 by a British-Pakistani student and the Supreme Court of Pakistan referred it to the government and the ECP for consideration \cite{pildatblog}. After a hiatus of almost two decades, more petitions followed in quick succession: in 2011, Dr. Arif Alvi, Secretary General of Pakistan Tehreek-e-Insaaf (PTI), a popular political party petitioned the Supreme court in this regard; in 2014, the Islamabad High Court was likewise petitioned by a concerned overseas citizen, and in June 2015, by the Chairman of PTI, Mr. Imran Khan.

The judiciary, while addressing this grievance, has upheld this fundamental right of overseas citizens on multiple occasions \cite{j2013} \cite{j2014} \cite{j2018}, ruling that this right cannot be denied on technical grounds, and it has repeatedly directed the ECP to make the necessary logistical arrangements. We discuss these attempts next.\\

\noindent \textbf{Efforts by the Election Commission and Parliament}
In 2012, in one of the earliest statements on the subject, the ECP dismissed the possibility of overseas citizens' participation in the General Elections of 2013, citing logistics and budgeting issues \cite{ecp2012}. However, by 2015, the ECP had established a Directorate for Overseas Voting in its Secretariat, which conducted mock overseas voting exercises using postal ballots and voting via telephone \cite{ecppilotreport}.

These trials were unsuccessful. The reasons were clarified in a study commissioned by the ECP: \emph{``We find that any remote voting solution using currently available technology — whether postal, internet, telephone, or proxy — will lack the necessary electoral integrity checks to preserve the credibility of an election result.''} Commenting on the feasibility of other modalities, the report stated: \emph{``...given the size and dispersal of the Pakistani diaspora, coupled with the limited official resources available in-country and abroad, any significant in-person voting operation would be expensive and logistically challenging''} \cite{undpfindings}.

In July of 2014, the Parliamentary Committee for Electoral Reforms was constituted with Finance Minister Ishaq Dar as chair. A sub-committee was formed in January, 2016 by MP Dr. Arif Alvi (one of the petitioners for overseas voting mentioned earlier and Secretary General of PTI) to devise a mechanism for overseas voting \cite{pcer} and in March, 2017, it proposed remote Internet voting as a potential solution. Consequently, the committee authored the Elections Act of 2017, which authorized the ECP \emph{``to conduct pilot projects for voting by Overseas Pakistanis in bye-elections''} \cite{pildatblog}. 

The ECP subsequently requested the National Database Registration Authority (NADRA) to build a system. NADRA is an independent and autonomous agency working under the Ministry of Interior and tasked with managing government databases and issuing national identity cards to citizens. However the system did not materialize: in June 2017, ECP again contacted NADRA, but NADRA expressed its regrets at not having a solution available \cite{ecppilotreport}.\\

\noindent \textbf{The Supreme Court Intervenes}
Interest in overseas voting peaked again in the six months leading up to the General Elections of July, 2018. The Supreme Court of Pakistan consolidated 16 similar constitutional petitions and resumed hearings on the issue \cite{16petitions}. It sought reports from the ECP and NADRA over non-compliance of Section 94 of the Elections Act (regarding pilot projects for overseas voters) \cite{cjpasks}. In an attempt to break the deadlock, the Supreme court directed NADRA to develop an Internet voting system. NADRA informed the court that it would require 4 months to build a system and would cost Rs.150 million (approximately US\$ 1.36 million) \cite{nadraquote}. However, the Court ordered NADRA to present it in 10 weeks \cite{10weeks}. 

The new system, i-Voting, was unveiled on April 12, 2018 at a public session convened by the Supreme Court \cite{invite}. The audience included members of various political parties, IT experts from Pakistani universities, concerned citizens, and members of the media. It was here that IT experts aired serious security concerns regarding this system and pointed out that similar systems had been demonstrably attacked and were being phased out in developed countries. The Supreme Court concluded the session by forming an Internet Voting Task Force (IVTF) to audit the system and assess its suitability for deployment in the forthcoming general elections of July, 2018.

 \section{The i-Voting System}
\label{sec:iVoting}

Here we describe NADRA's i-Voting system and summarize the IVTF findings.\\

\noindent \textbf{System Architecture} The i-Voting system conforms to the traditional design of Internet voting solutions, where a centralized database is used to store and tabulate votes, which voters access using a Web portal.

\begin{figure}[b]
\centering
\includegraphics[width=12cm]{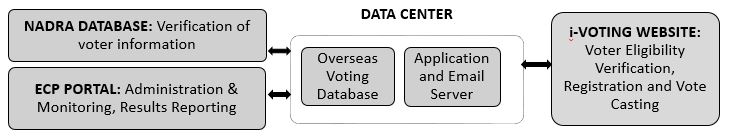}
\vspace{-2mm}
\caption{i-Voting: System Architecture}\label{fig:architecture}
\end{figure}

\begin{figure}
\centering
\includegraphics[width=9cm]{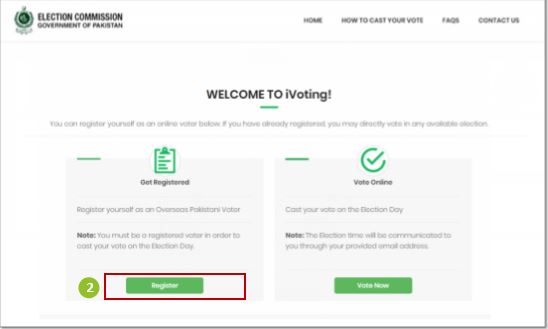}
\vspace{-3mm}
\caption{iVoting: Interface}\label{fig:interface}
\end{figure}

More specifically, as depicted in Fig.~\ref{fig:architecture}, a central datacenter hosts the overseas votes database and an application and email Server. These servers interface with a webserver hosting the i-Voting Web portal (Fig.~\ref{fig:interface}) and NADRA databases (for verification of voter information). The ECP can monitor the system using an administrative portal. Load balancing and backup arrangements are deployed as well as standard security solutions including firewalls, intrusion detection mechanisms, and mitigation of distributed denial of service (DDoS) attacks.

\begin{figure}[b]
\centering
\includegraphics[width=12cm]{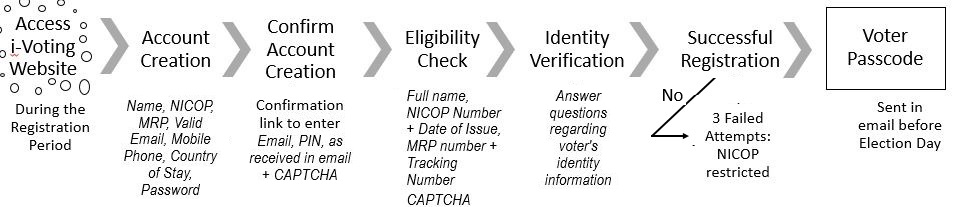}
\vspace{-2mm}
\caption{Voter Registration}\label{fig:registration}
\end{figure}

\noindent \textbf{Voter Registration} The registration process is depicted in Fig.~\ref{fig:registration} \cite{ivotingguide}. To enrol, a user must possess his/her passport, a National Identity Card for Overseas Pakistanis (NICOP), and a valid e-mail address. The ECP announces a public Registration Phase during which prospective voters enter their basic details into the system. A confirmation email with a PIN is then sent to the voter. To confirm the account, the user enters the PIN and solves a CAPTCHA.

Now the user logs in to the system and provides further details of his NICOP and passport. He also answers two randomly chosen questions pertaining to his identity after which he is successfully registered. The system allows a maximum of 3 answer attempts, failing which the NICOP number is restricted.



\noindent \textbf{Vote Casting and Preparation of Results}
Prior to polling day, each registered voter is emailed containing a unique passcode (which acts as a one-time password), enabling him to log on to his i-Voting account and cast a vote for his respective National Assembly and/or Provincial Assembly seat (Fig.~\ref{fig:registration}).


When polling concludes, the ECP tabulates the votes via the Reporting Portal and dispatches the tally to concerned officials for consolidation of results.\\

\begin{figure}[ht]
\centering
\includegraphics[width=12cm]{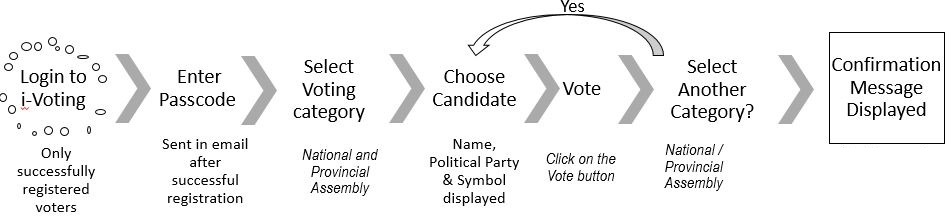}
\vspace{-2mm}
\caption{Vote Casting}\label{fig:voting}
\end{figure}
\vspace{-4mm}

\noindent \textbf{Internet Voting Task Force}
The Internet Voting Task Force (IVTF) was given a time window of 3 weeks to assess the security of the i-Voting system. The team comprised of IT and security specialists and academic researchers \cite{ecporder}. They conducted a high level security analysis of the system, examined the code, and mounted some typical attacks. The results were written up in a report and submitted to the Supreme Court. Their key findings are as follows:

\begin {enumerate}
\item i-Voting does not provide ballot secrecy, a fundamental right defined in the Constitution of Pakistan. This further opens up the possibility of vote buying and coercion of overseas voters.
\item A key security vulnerability allowed overseas voters to bypass their native constituencies and cast votes for any two seats of their choice in the country.
\item The IVTF successfully launched impersonation attacks, enabling them to send fake emails purportedly from the ECP to direct voters to fake websites.
\item  i-Voting avails the services of a leading DDoS mitigation solution, a measure which researchers have recently demonstrated can potentially compromise ballot secrecy and election integrity \cite{culnane}.
\item The system employs certain third-party security components (such as text-based CAPTCHAs) which are obsolete and demonstrably insecure.
\end {enumerate}

The IVTF also raised other critical non-security concerns: i-Voting lacked verifiability, fail-safe, or redundancy mechanisms. There were no security policies or procedural controls defined to protect critical security processes from insider attacks. No usability studies or trials had been conducted for the system. Futhermore, the system was built in an ad-hoc manner with key documentation missing. For instance, there was no documented Solution Requirements Specification (SRS) or documentation pertaining to key operational processes (such as administration, hosting, responsibility of critical components), which limited assessment for certain important security attacks.

The IVTF therefore strongly argued against the deployment of i-Voting in the upcoming General Elections of 2018. Their report stated that this would be \emph{``a hasty step with grave consequences''} The report also emphasized that \emph{``many of these security vulnerabilities are not specific to iVOTE [sic] but are inherent to this particular model of Internet voting systems''}  \cite{ivtfreport}.

The report also made various recommendations to facilitate overseas voters. We discuss these in Sec.~\ref{sec:discussion} and \ref{sec:wayforward}.

 \section{Deployment}
\label{sec:deployment}

The Supreme Court of Pakistan revisited the matter of Internet voting after the general elections in August, 2018, and ruled: \emph{``Based on these representations we prima facie find the mechanism of I-Voting (sic) to be safe, reliable and effective for being utilized in a pilot project. We are sanguine that the aforesaid proposed rules shall be incorporated in the Election Rules, 2017 to enable overseas Pakistanis to exercise their right of vote in the forthcoming bye-elections.''} The court further stipulated that votes cast using i-Voting not be added to the final tally until the ECP is satisfied with regards to their \emph{``technical efficacy, secrecy, and security''}. In case of any dispute the ECP was authorized to exclude these overseas votes from the official results \cite{j2018}.

The ECP consequently amended the Election Rules to accommodate the requirements of Internet voting. NADRA implemented certain technical recommendations of the IVTF\footnote{No details have been published on what specific changes were made.} and trained ECP officials to administer the system. The ECP launched a media campaign for voter awareness and published detailed guides and video tutorials for the i-Voting system. A dedicated support center was also set up to provide telephone and email assistance \cite{ecppilotreport}.\\

\noindent \textbf{First Pilot (Bye-Elections - 14 October, 2018)}
Bye-elections were held for 35 constituencies (11 National Assembly and 24 Provincial Assembly seats). The total overseas Pakistanis eligible to participate in these polls numbered a significant 631,909. However, out of these only 7,419 citizens (1.17\%) actually registered to vote using the new system. On the day of the elections, a total of 6,146 voters of these citizens cast their votes \cite{ecppilotreport}.

ECP later reported that on the day of the polls the system successfully withstood 7,476 DDoS attempts.\footnote{No details of these attacks have been released to the public.} The top 5 countries by voter count were the United Arab Emirates (1,654), Saudi Arabia (1,451), the United Kingdom (752), Canada (328), and the United States (298). The pilot project cost approximately Rs. 95 million (0.67 million USD approximately) \cite{ecppilotreport}.

The trial was smooth and uneventful. In its own report, the ECP attributed the low turnout to the short time frame within which the system was deployed and advertised. The ECP also cited key issues which echoed the concerns of the Internet Voting Task Force (discussed in Sec.~\ref{sec:iVoting}), in that the system violates ballot secrecy, enables voter coercion, lacks auditability, and may be vulnerable to state-level cyberattacks.\\

\noindent \textbf{Second Pilot (Bye-Elections - 13 December, 2018)}
As many as 4,667 overseas Pakistanis from more than ten countries were eligible to vote for one Provincial Assembly seat \cite{fafenivoting}. However, only 77 overseas Pakistanis registered to vote. The ECP has not released any further details about this trial \cite{pp168}.
 \section {Materiality - Are Overseas Votes Decisive?}
\label{sec:statistical}

In this section we undertake a basic post-election analysis to examine the potential impact of overseas votes on final results.

The leading Pakistani citizen observation group, Free and Fair Election Network (FAFEN) has conducted an analysis of 2018 General Election results \cite{fafen_MoV} and has determined that, in a significant number of constituencies, the Margin of Victory (MoV) is less than the number of invalid votes. A similar analysis, but comparing Margin of Victory with number of Eligible Overseas Voters would be useful to highlight the materiality\footnote{Materiality in this context refers to the theoretical scenario where all possible overseas votes are cast, and all are cast for the second place or losing candidate, the outcome of the election might have been different/might be different.}.
There is no publicly available voter registration or population data which shows how many registered voters are actually Overseas Pakistanis for all constituencies. The only exception to this is for the constituencies where by elections were conducted using i-Voting in October and December 2018. The average percentage of eligible overseas voters, as a percentage of total voters in the October 2018 By-elections is 6.88\% ( Table 2). With this assumption, we estimate the number for eligible overseas voters for individual constituencies under scrutiny. 

We then calculate an estimated Overseas Pakistani Voters value for each contested constituency in the October 2018 By elections that was also contested in the July 2018 General Elections. We now compare these Estimated Overseas Voters (EOV) value with the margin of victory and flag where the MoV is less. We do this both for the October 2018 By-elections and the July 2018 General Elections.
We may describe the number of Overseas Pakistani Voters in these cases as material to the outcome of the election\footnote{The public domain sources for  Table 2 and 3 are no longer available on ECP Website. The documents will be made available at a URL which will be cited later (to retain anonymity).}.

As Table \ref{tab:stats2} shows, in ten of twenty-seven by-election races, Overseas Pakistani Voters had the potential to be material. This grows to thirteen of twenty-seven for General Election races. In five races, both Bye-Election and General Election saw Margin of Victory less than estimated Overseas Pakistani Voters. It is, we believe, reasonable to assume that, in competitive future General Elections at least one in five races may be decided by votes cast by overseas Pakistani voters. This places the integrity of and trust in any internet voting solutions deployed by the Elections Commission of Pakistan into very sharp focus. This is positive in the sense that overseas Pakistanis can feel their votes count. At the same time it necessitates election integrity checks so that this right is not misused. 

\vspace{5mm}
\begin{table*}[t]
{\tiny
  \centering
    \begin{tabular}{cc}
    \toprule
    \textbf{Description} & \textbf{Number}    \\
    \midrule
    Total Registered Voters in bye-election  &      9,185,705 \cite{}  \\
    \midrule
    Total Eligible NICOP &         734,777 \cite{}   \\
    \midrule
    Non-Machine Readable Passports ( 14\%) &         102,868 \cite{}  \\
    \midrule
    Estimated Total Overseas Pakistani Voters &  631,909 \cite{} \\
    \midrule
    Eligible Overseas Pakistani Voters as \% of Voters Registered & 6.88\% \cite{} \\
        \bottomrule
  \end{tabular}
    \label{tab:stats1}
    \vspace{2mm}
    \caption{Calculating Estimated Percentage of Overseas Pakistani Voters}}
\end{table*}

\begin{table*}[t]
{\tiny
  \centering
    \begin{tabular}{cccccccc}
    \textbf{Constit} & \textbf{ Total } & \textbf{ Estimated  } & \multicolumn{2}{c}{\textbf{ General Elections }} & \multicolumn{2}{c}{\textbf{Bye Elections}} & \textbf{ Both } \\
    \textbf{uency} & \textbf{ Registered } & \textbf{ Overseas } & \multicolumn{2}{c}{\textbf{ July,2018 }} & \multicolumn{2}{c}{\textbf{October,2018}} & \textbf{ Elections } \\
          & \textbf{ Voters } & \textbf{ Voters (EOV) } & \textbf{ MoV } & \textbf{MoV \textless EOV} & \textbf{ MoV } & \textbf{MoV \textless EOV} & \textbf{ Mov \textless EOV } \\
    \midrule
    PB-40 &      76,173  &             5,240  &    13,345  & -     &     9,141  & -     & - \\
    \midrule
    NA-53 &    313,141  &           21,541  &    48,763  & -     &   18,630  & -     & - \\
    \midrule
    NA-35 &    582,785  &           40,091  &      7,001  & Yes   &   23,455  & -     & - \\
    \midrule
    PK-3  &    146,180  &           10,056  &      5,550  & Yes   &     1,163  & -     & - \\
    \midrule
    PK-7  &    155,719  &           10,712  &      5,825  & Yes   &        334  & -     & - \\
    \midrule
    PK-44 &    202,601  &           13,937  &    10,857  & Yes   &     1,630  & -     & - \\
    \midrule
    PK-53 &    153,352  &           10,549  &      6,729  & Yes   &          61  & -     & - \\
    \midrule
    PK-61 &    139,517  &             9,597  &      4,593  & Yes   &     5,247  & -     & - \\
    \midrule
    PK-64 &    160,728  &           11,056  &    18,579  & -     &   13,215  & -     & - \\
    \midrule
    PK-97 &    155,032  &           10,665  &    16,461  & -     &   10,172  & -     & - \\
    \midrule
    NA-56 &    640,133  &           44,036  &    64,490  & -     &   41,593  & Yes   & - \\
    \midrule
    NA-63 &    371,713  &           25,571  &    35,979  & -     &   26,292  & -     & - \\
    \midrule
    NA-65 &    553,289  &           38,062  &    51,963  & -     &   68,591  & -     & - \\
    \midrule
    NA-69 &    469,177  &           32,275  &    73,172  & -     &   50,803  & -     & - \\
    \midrule
    NA-124 &    535,172  &           36,815  &    65,287  & -     &   47,533  & -     & - \\
    \midrule
    NA-131 &    365,677  &           25,155  &         756  & Yes   &   10,031  & Yes   & Yes \\
    \midrule
    PP-3  &    224,755  &           15,461  &    37,008  & -     &        227  & Yes   & - \\
    \midrule
    PP-27 &    312,370  &           21,488  &      1,766  & Yes   &        656  & Yes   & Yes \\
    \midrule
    PP-118 &    222,190  &           15,285  &         548  & Yes   &     5,189  & Yes   & Yes \\
    \midrule
    PP-164 &    137,906  &             9,486  &    20,870  & -     &     7,561  & Yes   & - \\
    \midrule
    PP-165 &    132,077  &             9,085  &    20,372  & -     &     5,742  & Yes   & - \\
    \midrule
    PP-201 &    232,120  &           15,968  &    17,297  & -     &     7,024  & Yes   & - \\
    \midrule
    PP-222 &    196,858  &           13,542  &    11,446  & Yes   &     6,083  & Yes   & Yes \\
    \midrule
    PP-261 &    187,510  &           12,899  &      9,371  & Yes   &   14,261  & -     & - \\
    \midrule
    PP-272 &    167,467  &           11,520  &      5,390  & Yes   &     8,899  & Yes   & Yes \\
    \midrule
    PP-292 &    141,297  &             9,720  &         253  & Yes   &   10,692  & -     & - \\
    \midrule
    NA-243 &    402,731  &           27,704  &    67,291  & -     &   21,601  & -     & - \\
    \end{tabular}%
    }
    \caption{Materiality of Overseas Voters in Selected Constituencies, General and Bye-Elections 2018}
  \label{tab:stats2}%
\end{table*}%

 \section{Discussion}
\label{sec:discussion}
In this section we further examine outstanding issues arising from this experiment and we attempt to contextualize these by examining the unique political and institutional factors that motivated these pilot projects.

\subsection{Ballot Secrecy and Voter Coercion}
In delivering one right (the right of overseas Pakistanis to vote), the solution risks undermining another (the right to secrecy). The i-Voting system does not comply with Article 226 of the Constitution of Pakistan \cite{constitution226} and the Elections Act 2017, Section 81\footnote{The exceptions in Section 81 are not to the secrecy requirement, rather to requirement of casting a vote by inserting paper ballot into a ballot box.} \cite{electionsact2017}, that impose ballot secrecy. Being a remote voting modality, there is no mechanism to prevent an individual from revealing their vote to others. Similarly, certain event logging software (specially on a shared/public device) can secretly capture the choice of a voter. 

\begin{figure}[t]
\centering
\includegraphics[width=9cm]{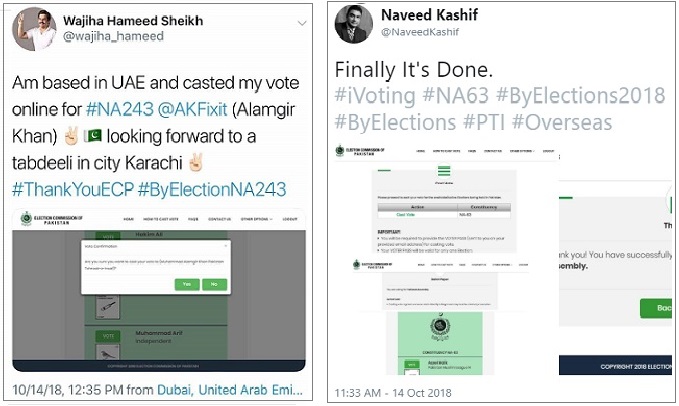}
\caption{Twitter users Posting Screenshot of Vote}\label{fig:offence}
\end{figure}

Electoral offences were committed by voters unintentionally, by posting screen shots on social media, as shown in Fig ~\ref{fig:offence}. The tweeters seem to be unaware their actions are electoral offences, and being outside the jurisdiction of Pakistan it is unclear, how such offenders can be brought to justice\footnote{Section 178 of the Elections Act 2017 elaborates the offences relating to ballot secrecy.}.

In addition, to the lack of secrecy for the voter at the client end, the low levels of participation in pilots also mean that, in some cases, typically PA seats (PB-35, PP-165, PP-292 from October 2018 bye-elections \cite{ecppilotreport}), the voter's choice is revealed. The usual solution to this problem (mixing votes from multiple ballot boxes or polling stations) is not available in the i-Voting context or could only be implemented at the cost of further erosion of already minimal transparency. 

As the IVTF report points out, some jurisdictions \cite{secrecy} allow a voter to waive their right to secrecy. This is not a solution to the problem, as any voter (or party or candidate) to assert their constitutional and legal rights to secrecy for the system to be challenged. 

Almost half of the diaspora, over 4 million Pakistanis reside in the Middle East, and about a quarter (over 1.5 million)  reside in Europe \cite{opp}. A bulk of the diaspora specifically in the Middle East are labourers. The ECP itself recognizes the risk of vote buying and coercion when it speaks of the "kafeel"\footnote{Sponsor for a migrant worker} abusing custody of passports \cite{ecppilotreport}. The ILO\footnote{International Labour Organization} describes this system as placing migrant workers in \emph{''a position of vulnerability and have very little leverage to negotiate with employers, given the significant power imbalance embedded within the employment relationship. Common grievances expressed by migrant workers include restrictions on free movement, confiscation of passports, delayed or non-payment of salaries, long working hours, untreated medical needs, and violence – all conditions that can give rise to situations of forced labour and human trafficking''} \cite{ilomigrant}. It is reasonable to assume that anyone who will treat migrant workers in this manner will not hesitate to exploit their votes for political or financial benefit.

Migrant workers are bound to face difficulty to independently use the i-Voting System. This could pave the way for coercion, vote buying  and compromise secrecy if vote casting is aided by a computer literate party. Thus, usability tests need to be conducted to receive direct input from real users. It might be argued that low usability, was a primary reason of the low registration turnout, where only 1.17 \% \cite{ecppilotreport} of the total eligible overseas voters successfully registered with the i-Voting system. Further, there seem to be no special accessibility features incorporated to address the needs of voters with disabilities.

\subsection{Voter Authentication}
The process of registration on the i-Voting platform \cite{ivotingwebsite} is entirely out of ECP's control, relying as it does on a verification method conducted and adjudicated by a computer programme. Potential overseas voters are quizzed with questions, whose answers are considered \emph{''secret''}. Common sense dictates that, despite the familial/personal nature of these questions, the answers will not be known exclusively to the voter. As a consequence, ECP cannot guarantee that the voter registered via the online platform \cite{ivotingwebsite} is indeed the eligible citizen, or an impostor. Other key mechanisms to protect the integrity of the electoral rolls - the public display of and claims/objections on the draft electoral rolls - are omitted from the online process. Political parties, observers, and voters themselves, are not given the access to these electoral rolls to allow for the scrutiny that would contribute to stakeholder confidence in the electoral rolls.

Furthermore, the mechanism within i-Voting to \emph{``lock''} an identity following repeated incorrect answers or CAPTCHA verification may be used for voter suppression - a sort of denial-of-service attack, albeit on a vote-by-vote basis. Voters may not know the answers to all the questions they might be asked (where, for example, an 18 year old was registered and a parent provided the information). Corrupt or partisan Presiding Officers could merely strike out the names of legitimate voters saying that they had registered online for i-Voting.

\subsection{Election Integrity and Dispute Resolution}
On election day in polling stations across Pakistan, a long list of integrity mechanisms are in place, arising from the Constitution, the Elections Act 2017 and the Election Rules 2017, as amended\footnote{The election is conducted in full view of polling staff, party/candidate agents and observers, who first-hand witness the integrity checks in place: verify ballot boxes are initially empty, identify voters on arrival, ink their fingers (to prevent multiple voting), vote casting in secrecy behind a screen, placing the ballot paper into the transparent ballot box, the Presiding officer conducting the count and disseminating the results form to all stakeholders, and packaging all ballot papers (valid, invalid, challenged, spoiled) separately in tamper-evident envelopes.}.  In the i-Voting system these fourteen separate mechanisms are missing with consequences for electoral integrity. The exclusion of party/candidate agents and citizen observers from the i-Voting process is compounded by the inherent absence of any verifiability mechanism or possibility to audit the i-Voting system - by design -\emph{''In order to ensure that Voting is kept secret, all data was encrypted and no audit trail of voting was kept by the system''} \cite{ecppilotreport}.

 ECP may exclude based on its \emph{''opinion''} as to whether the \emph{''technical efficacy, secrecy and security of the voting has not been maintained''} \cite{ecppilotreport}. It is not known how ECP informs that opinion, or whether it has the required access to the i-Voting system. Given the 2018 recourse to establishing the IVTF, it seems likely that ECP lacks the technical capacity to properly arrive at an informed opinion. Given the likely materiality of votes cast by overseas Pakistanis in a significant proportion of contests, we may expect many electoral disputes to centre around the integrity of i-Voting system. 

Specifically, in a developing country like Pakistan, where the democratic process is at an inflection point, and the mechanisms to investigate and resolve electoral disputes,  are still very fragile, electoral improprieties or even the impression of such can potentially lead to political deadlock and turmoil. An indication of this is the PILDAT (Pakistan Institute of Legislative Development and Transparency) report on the perception of pre-poll fairness which notes that \emph{''the internet-based OP voting may also be a major instrument of rigging in 2018 General Election''} \cite{pildatprepoll}. Diverse stakeholders including the ECP itself \cite{ecppilotreport}, and prominent mainstream political parties expressed similar reservations \cite{pmln} \cite{ppp}. Lack of auditability features means there is no evidence if results are challenged through an election petition. These, coupled with questions over the capacity and  willingness of the judiciary, raises concerns about the resolution of electoral disputes around internet voting.

\subsection{Threat Model}
Concerns have also been raised regarding the threat model on which i-Voting is based. The recent controversy of foreign interference in US elections hints that a developing country like Pakistan may also be at risk. ECP's report on the pilot deployment highlights this concern: \emph{``[adversaries] did not materially interfere merely to put us off track. When the system is finalized and put into practice in the next elections, shall we be able to counter/control cyber attacks[?]''} \cite{ecppilotreport}. 
    
Furthermore, whereas the high number of DDoS attacks (around 7476 on bye-elections day), posed an outage threat, the use of a DDoS mitigation service, as demonstrated recently by Culnane et al. \cite{culnane}, introduces a new attack vector. The mitigation service is in a position to decrypt incoming traffic, thereby able to compromise ballot secrecy and potentially even alter the content. The IVTF audit highlighted this concern in their report and pointed out that the servers employed by the DDoS mitigation service were all based overseas and beyond control of Pakistani authorities.

\subsection{The Curious Case of Pakistan}

We see the Supreme Court at the forefront, driving the institutions to deliver a voting solution to overseas Pakistanis.  Here we try to make sense of the unique predicament and examine the various factors that led to this situation. The judiciary has time and again reiterated the ECP to roll out a voting mechanism for overseas Pakistanis, but to no avail. A concrete step in this direction was long overdue and it had to take the Supreme Court to push it through, given the institutional inertia within the government.

The Supreme Court of Pakistan frequently takes Government ministries and other public bodies to task for not fulfilling their obligations \cite{cdapromise}. Whether through judicial activism (using \emph{suo moto} powers) or responding to petitions from interested parties, it often gets involved in the technical specifics of cases. Its jurisdiction \emph{``is not limited to mere procedural technicalities as it enjoys certain inherent powers to do complete justice in any case''} \cite{shabbir}. A vivid example of the Supreme Court's ambition beyond procedural technicalities is the fund established in July 2018 to raise money to build dams. This fund currently exceeds ten billion Pakistani rupees (approximately 71 million US \$) \cite{scopdamfund}.

Cognizant of this deviation, the Honourable Chief Justice of Pakistan, inquired whether it was the job of the Supreme Court to give the right to vote to overseas Pakistanis? \cite{cjpasks} In the matter of how best to enfranchise overseas Pakistanis the Supreme Court initially directed the elections management body to develop an internet voting system and then later mandated the use of this system in binding political bye-elections. In doing so, the Supreme Court dismissed unambiguous and dire warnings from the IVTF about the hazards of the proposed system as mere \emph{''technical and security apprehensions''}, and that the report was \emph{''generally positive and encouraging''} \cite{j2018}. While the IVTF report clearly says \emph{''Hopefully, this discussion thus far demonstrates to the reader why internet voting is recognized by security experts to be a controversial and risky undertaking''}, and it concludes by asserting \emph{''We would, therefore, urge all stakeholders to exercise extreme caution in approaching the question of internet voting''} \cite{ivtfreport}. This disconnect has received media recognition \cite{riskyvoting}. 

The ECP itself was very reluctant to adopt this modality of overseas voting. Recently, when the matter was taken up in the Senate, in May 2019, Senator Javed Abbasi recollected that \emph{''the ECP had convinced political parties that the system should not be introduced in Pakistan, but could not convince the Supreme Court''}, at which an ECP representative expressed his dismay that while the ECP tried to dissuade the Supreme Court, no political party supported the ECP in Supreme Court \cite{ecpsenate}. The absence of a broad political consensus on the use of i-Voting to enfranchise the diaspora does not bode well for the future.

Neither the Elections Commission of Pakistan, nor the developers of the i-Voting system challenged the Supreme Court's interpretation of the IVTF findings or recommendations. Since there is no higher court than the Supreme Court, no appeal is possible. If a future election is decided on votes cast by overseas Pakistanis, via the i-Voting system, and the result is challenged  \textemdash  which seems highly likely, given both the deficiencies and materiality described earlier in this paper \textemdash  it will be interesting to see if the electoral dispute resolution process ends up in the same court. \section{Way Forward and Conclusion}
\label{sec:wayforward}

Pakistan's experiment for the October 2018 by elections was the largest deployment of Internet voting in a binding election, anywhere in the world.  The recommendations of both the IVTF report, and ECP's own report on the October 2018 pilot exercise are comprehensive and we endorse these.  Going beyond these, and comparing the Pakistani experience with other countries who are further along the internet voting pathway, we would highlight two vital priorities. First, transparency: ECP and NADRA succeeded in delivering a working prototype system in the short time available, but the details of the process were, and remain, opaque. Stakeholder acceptance cannot be assured in future without meaningful transparency and greater consultation, having regard to voter secrecy. Second, capacity building across all stakeholders, starting with, and led by ECP (such as establishing a dedicated R\&D cell within the ECP), to deliver competent national ownership and informed policymaking. It seems likely that escalation from pilots in bye-elections to full-scale use of internet voting for the enormous Pakistani diaspora will happen in 2023. The issues highlighted in this paper should receive urgent attention by all Pakistani stakeholders.


\end{document}